\documentclass[conference]{IEEEtran}
\ifCLASSINFOpdf
  \usepackage[pdftex]{graphicx}
   \graphicspath{{figs/}}
\else
\fi
\usepackage{float}
\usepackage{framed}
\usepackage[utf8]{inputenc}
\usepackage[T1]{fontenc}
\usepackage{microtype}
\usepackage{graphicx}
\usepackage{dblfloatfix}
\usepackage[noadjust]{cite}
\usepackage[hidelinks, pdftitle={Exploring Explainability: A Definition, a Model, and a Knowledge Catalogue},
            pdfsubject={Requirements Engineering, Software Engineering},
            pdfauthor={Larissa Chazette, Wasja Brunotte, Timo Speith},
            pdfkeywords={Explainability; Explanations; Explainable Artificial Intelligence; Interpretability; Non-Functional Requirements; Quality Aspects; Requirements Synergy; Software Transparency}]{hyperref}
\usepackage[keeplastbox]{flushend}
\usepackage{amsmath}
\usepackage{url}
\usepackage{enumitem}

\usepackage{csquotes}
\usepackage{color}

\usepackage{scalerel}
\usepackage{tikz}
\usetikzlibrary{svg.path}

\newcommand{\denquote}[1]{\textquoteleft#1\textquoteright}

\definecolor{orcidlogocol}{HTML}{A6CE39}
\tikzset{
  orcidlogo/.pic={
    \fill[orcidlogocol] svg{M256,128c0,70.7-57.3,128-128,128C57.3,256,0,198.7,0,128C0,57.3,57.3,0,128,0C198.7,0,256,57.3,256,128z};
    \fill[white] svg{M86.3,186.2H70.9V79.1h15.4v48.4V186.2z}
                 svg{M108.9,79.1h41.6c39.6,0,57,28.3,57,53.6c0,27.5-21.5,53.6-56.8,53.6h-41.8V79.1z M124.3,172.4h24.5c34.9,0,42.9-26.5,42.9-39.7c0-21.5-13.7-39.7-43.7-39.7h-23.7V172.4z}
                 svg{M88.7,56.8c0,5.5-4.5,10.1-10.1,10.1c-5.6,0-10.1-4.6-10.1-10.1c0-5.6,4.5-10.1,10.1-10.1C84.2,46.7,88.7,51.3,88.7,56.8z};
  }
}

\newcommand\orcidicon[1]{\href{https://orcid.org/#1}{\mbox{\scalerel*{
\begin{tikzpicture}[yscale=-1,transform shape]
\pic{orcidlogo};
\end{tikzpicture}
}{|}}}}

\newcommand\copyrighttext{%
  \footnotesize \textcopyright~2021 IEEE. Personal use of this material is permitted.
  Permission from IEEE must be obtained for all other uses, in any current or future
  media, including reprinting/republishing this material for advertising or promotional
  purposes, creating new collective works, for resale or redistribution to servers or
  lists, or reuse of any copyrighted component of this work in other works.
  }
\newcommand\copyrightnotice{%
\begin{tikzpicture}[remember picture,overlay]
\node[anchor=south,yshift=10pt] at (current page.south) {\fbox{\parbox{\dimexpr\textwidth-\fboxsep-\fboxrule\relax}{\copyrighttext}}};
\end{tikzpicture}%
}

\begin{document}
%\title{Engineering Explainable Systems:\\The Impact of Explainability on System Quality}
%\title{The Impact of Explainability on System Quality: \\Modeling a Knowledge Catalogue}
%\title{A Catalogue of Explainability:\\The Impact of Explainability on System Quality}
%The Impact of Explainability on System Quality - A Catalogue
%A Catalogue of Explainability:\\A Model of Dimensions and Impacts
%A Catalogue for Explainability: Synergies and Conflicts
%A Knowledge Catalogue of Explainability:  Modelling its Impact on System Quality
%A Knowledge Catalogue of Explainability: Understanding The Various Dimensions
%\title{A Knowledge Catalogue of Explainability: Diving Into Its Dimensions}
%\title{A Knowledge Catalogue for Explainability: Definitions, Impacts, and Dimensions}

%\title{How Does Explainability Impact a System? \\ A Knowledge Catalogue}
%\title{Exploring Explainability: A Definition, a Model, and a Knowledge Catalogue}
\title{Exploring Explainability: \\ A Definition, a Model, and a Knowledge Catalogue}

% author names and affiliation
% \author{\IEEEauthorblockN{Larissa Chazette\IEEEauthorrefmark{1}\orcidicon{0000-0001-6093-8875}, Wasja Brunotte\IEEEauthorrefmark{1}\IEEEauthorrefmark{2}\orcidicon{0000-0003-4127-6508}, Timo Speith\IEEEauthorrefmark{3}\orcidicon{0000-0002-6675-154X}}

% \IEEEauthorblockA{\IEEEauthorrefmark{1}Leibniz University Hannover, Software Engineering Group, Hannover, Germany\\}
% \IEEEauthorblockA{\IEEEauthorrefmark{2}Leibniz University Hannover, Cluster of Excellence PhoenixD, Hannover, Germany\\}
% \IEEEauthorblockA{\IEEEauthorrefmark{3}Saarland University, Institute of Philosophy and Department of Computer Science, Saarbrücken, Germany\\
% 	Email: \{larissa.chazette, wasja.brunotte\}@inf.uni-hannover.de, timo.speith@uni-saarland.de}
% }

% Ohne ORCID
\author{\IEEEauthorblockN{Larissa Chazette\IEEEauthorrefmark{1}, Wasja Brunotte\IEEEauthorrefmark{1}\IEEEauthorrefmark{2}\, Timo Speith\IEEEauthorrefmark{3}}

\IEEEauthorblockA{\IEEEauthorrefmark{1}Leibniz University Hannover, Software Engineering Group, Hannover, Germany\\}
\IEEEauthorblockA{\IEEEauthorrefmark{2}Leibniz University Hannover, Cluster of Excellence PhoenixD, Hannover, Germany\\}
\IEEEauthorblockA{\IEEEauthorrefmark{3}Saarland University, Institute of Philosophy and Department of Computer Science, Saarbrücken, Germany\\
	Email: \{larissa.chazette, wasja.brunotte\}@inf.uni-hannover.de, timo.speith@uni-saarland.de}
}

% DOUBLE-BLIND-READY
%\author{\IEEEauthorblockN{Placeholder for the authors' names}
%\IEEEauthorblockA{Placeholder for additional information\\}
%\IEEEauthorblockA{Placeholder for additional information\\
%	Placeholder fr additional information}
%}

\nocite{AdditionalMaterial}
%
%% conference papers do not typically use \thanks and this command

%Telephone: (800) 555--1212, Fax: (888) 555--1212}


\maketitle

\copyrightnotice
\vspace{-2ex}

% \fi
\IEEEpeerreviewmaketitle

\begin{abstract}
The growing complexity of software systems and the influence of software-supported decisions in our society awoke the need for software that is transparent, accountable, and trustworthy. Explainability has been identified as a means to achieve these qualities. It is recognized as an emerging non-functional requirement (NFR) that has a significant impact on system quality. However, in order to incorporate this NFR into systems, we need to understand \emph{what} explainability means from a software engineering perspective and \emph{how} it impacts other quality aspects in a system. This allows for an early analysis of the benefits and possible design issues that arise from interrelationships between different quality aspects. Nevertheless, explainability is currently under-researched in the domain of requirements engineering and there is a lack of conceptual models and knowledge catalogues that support the requirements engineering process and system design. In this work, we bridge this gap by proposing a definition, a model, and a catalogue for explainability. They illustrate how explainability interacts with other quality aspects and how it may impact various quality dimensions of a system. To this end, we conducted an interdisciplinary Systematic Literature Review and validated our findings with experts in workshops.
\end{abstract}

\section{Introduction}

We live in the age of artificial intelligence (AI). Software decision-making has spread from simple daily decisions, such as the choice of a navigation route, to more critical ones, such as the diagnosis of cancer patients \cite{Panesar2019}. Systems have been strongly influencing various aspects of our lives with their outputs but can be as mysterious as black boxes to us \cite{Lepri2018}. This ubiquitous influence of black-box systems has induced discussions about the transparency and ethics of modern systems \cite{Adadi2018}. Responsible collection and use of data, privacy, and safety are just a few among many concerns. It is crucial to understand how to incorporate these concerns into systems and, thus, how to deal with them during requirements engineering (RE). 

Explainability is increasingly seen as the preferred solution to mitigate a system's lack of transparency and should be treated as a non-functional requirement (NFR) \cite{Koehl2019}. Incorporating explainability can mitigate software opacity, thereby helping users understand why a system produced a particular result and supporting them in making better decisions. Explainability also has an impact on the relationship of trust in and reliance on a system \cite{Bussone2015}, and it may avoid feelings of frustration \cite{Winkler2017}. Although explainability has been identified as an essential NFR for software-supported decisions \cite{Abdollahi2018} and one of the pillars for trustworthy AI \cite{Thiebes2020}, there is still a lack of an extensive overview that investigates the impact of explainability on a system.

In this paper, we investigate the concept of explainability and its interaction with other quality aspects. We use the notion of \emph{quality aspects} to refer both to \emph{NFRs} and to \emph{aspects} that relate to or compose NFRs. In this regard, we follow Glinz and see NFRs as attributes of or constraints on a system \cite{Glinz2007b}.  

Previous studies have shown that explainability is not only a means of achieving transparency and building trust but that it is also linked to other important NFRs, such as usability and auditability \cite{Chazette2020, Hind2019, Rosenfeld2019}. Explainability can, however, have both a positive and a negative impact on a system. Like other NFRs, explainability is difficult to elicit, negotiate, and validate. Eliciting and modeling NFRs is often a challenge for requirements engineers due to the subjective, interactive, and relative nature of NFRs \cite{Mairiza2011}. Often there are trade-offs between NFRs in a system that must be identified and resolved during the requirements analysis \cite{Mairiza2011, Cysneiros07}. 

%One of the challenges with respect to NFRs stems from the fact that information concerning them is rather tacit, distributed, and based on experience \cite{Chung2012},~\cite{gacitua2009making}. To mitigate this, a usual strategy adopted by requirements engineers to deal with NFRs during RE is to use pre-established lists or knowledge catalogues \cite{Mairiza2011}. These catalogues may compile knowledge about specific NFRs and their interactions with other quality aspects. Among others, catalogues support the elicitation process, where requirements engineers can use them to ask stakeholders about their interest with respect to the general system quality \cite{Portugal2018}. Catalogues also support the analysis of trade-offs where it is important to understand how two or more NFRs will interact in a system and how they can coexist (e.g., usability and security) \cite{gutmann2005security}. Existing works propose to build such catalogues to capture and structure knowledge that is scattered among several sources  \cite{Mairiza2011}, \cite{Serrano2013}, \cite{Carvalho2020}. 

One of the challenges with respect to NFRs stems from the fact that information concerning them is rather tacit, distributed, and based on experience \cite{Chung2012, Gacitua2009}. To mitigate this, a usual strategy adopted by requirements engineers to deal with NFRs during RE is to make use of artifacts such as conceptual models \cite{Chung2012}, pre-established lists, or knowledge catalogues \cite{Mairiza2011}. Models and catalogues can be used to help specify quality requirements \cite{Deissenboeck2009}. They may compile knowledge about specific NFRs and their interactions with other quality aspects. Among others, such artifacts support the elicitation process. Models can be used to understand the taxonomy of a given quality aspect during analysis, while catalogues can support the analysis of trade-offs where it is essential to understand how two or more NFRs will interact in a system and how they can coexist \cite{Gutmann2005}. Requirements engineers can also use models and catalogues to ask stakeholders about their interest with respect to the general system quality \cite{Portugal2018}. Existing works propose to build such artifacts to capture and structure knowledge that is scattered among several sources \cite{Mairiza2011, Chung2012, Serrano2013, Carvalho2020}. 

Since explainability is an emerging requirement, there is still a lack of structured knowledge about this NFR. To bridge this gap, we employed a multi-method research strategy consisting of an interdisciplinary Systematic Literature Review (SLR) and workshops. Overall, our goal is to advance the knowledge towards a common terminology and semantics, facilitating the discussion and analysis of explainability during the RE process. To this end, we distill definitions of explainability into an own suggestion that is useful for software and requirements engineering. We use this definition as a starting point to create a model that represents the impacts of explainability across different quality dimensions. Finally, we construct a knowledge catalogue of explainability and its impacts that is framed along these dimensions. %We combined the SLR with a method of qualitative data analysis and validated the findings by conducting two workshops.

%This paper is structured as follows: in the following section we present background and related work. In Sec. \ref{sec:ResearchDesign}, we raise our research questions (RQs), and outline the chosen research design. In Sec. \ref{sec:Definition}, we suggest our definition of explainability and in Sec. \ref{sec:Model}, we introduce our model. In Sec. \ref{sec:Catalogue}, we present the explainability catalogue and in Sec. \ref{sec:Discussion}, we discuss our results. Threats to validity are debated in Sec. \ref{sec:ThreatsToValidity}. Finally, we conclude our paper in Sec. \ref{sec:Conclusion}.
\section{Background and Related Work}
\label{sec:BackgroundAndRW}
Chung et al. \cite{Chung2012} explain the importance of conceptual models and knowledge catalogues as resources for the use and re-use of knowledge during system development. Models and catalogues can compile either abstract or concrete knowledge. At a more abstract level, such artifacts can compile knowledge about different NFRs and their interrelationships with other quality aspects. Likewise, models and catalogues can also compile more concrete knowledge, such as about methods and techniques in the field that can be used to operationalize a given NFR. The knowledge required to develop such artifacts is typically derived from literature, previous experiences, and domain expertise. By making this knowledge available in a single framework, developers can draw on know-how beyond their own fields and use this knowledge to meet the needs of a particular project. Essentially, software engineers can use models and knowledge catalogues to facilitate the software design process. %In this work, we built a conceptual model to frame our catalogue. %Since the knowledge in the resulting catalogue is the main source of distilled knowledge, we focus on related work on knowledge catalogues in this section.

Some researchers developed catalogues for specific domains based on the premise of the NFR framework. Serrano and Serrano \cite{Serrano2013} developed a catalogue specifically for the ubiquitous, pervasive, and mobile computing domain. Torres and Martins \cite{Torres2018} propose the use of NFR catalogues in the construction of RFID middleware applications to alleviate the challenges of NFR elicitation in autonomous systems. They argue that the use of catalogues can reduce or even eliminate possible faults in the identification of functional and non-functional requirements. Carvalho et al.~\cite{Carvalho2020b} propose a catalogue for invisibility requirements focused on the domain of ubiquitous computing applications. They emphasize the importance of software engineers understanding the relationships between requirements in order to select appropriate strategies to satisfy invisibility and traditional NFRs. Furthermore, they discovered that invisibility may impact other essential NFRs for the domain, such as usability, security and reliability.

On a general level, Mairiza et al.~\cite{Mairiza2011} conducted a literature review to identify conflicts among existing NFRs. They constructed a catalogue to synthesize the results and suggest that it can assist software developers in identifying, analyzing, and resolving conflicts between NFRs. Carvalho et al.~\cite{Carvalho2020} identified 102 NFR catalogues in the literature after conducting a systematic mapping study. They found that the most frequently cited NFRs were performance, security, usability, and reliability. Furthermore, they found that the catalogues are represented in different ways, such as softgoal interdependency graphs, matrices, and tables. The existence of so many catalogues illustrates their importance for RE and software design. Although these catalogues present knowledge about 86 different NFRs, none of them addresses explainability.

Since explainability has rapidly expanded as a research field in the last years, publications about this topic have become quite numerous, and it is hard to keep track of the terms, methods, and results that came up. For this reason, there have been numerous SLRs presenting overviews concerning certain aspects (e.g., used methods or definitions) of explainability research. Many of these reviews focus on a specific community or application domain. For instance, \cite{Nunes2017} focuses on explainability of recommender systems, \cite{Anjomshoae2019} on explainability of robots and human-robot interaction, \cite{Abdul2018} on the human-computer interaction (HCI) domain, and \cite{Mathews2019} on biomedical and malware classification. Another focus of these reviews is to demarcate different, but related terms often used in explainability research (e.g., in \cite{Adadi2018} and \cite{Arrieta2020}). For instance, the terms "explainablilty" and "interpretability" are sometimes used as synonyms and sometimes not.

Our review differs from others in the following ways. First, many other reviews do not have an interdisciplinary focus. Even if they do not focus on a specific community, they rarely incorporate views on explainability outside of computer science. Second, quality aspects are the pivotal focus of our work. To the best of our knowledge, only a few reviews explicitly include NFRs or quality aspects (most notably \cite{Nunes2017} and \cite{Langer2021}). Finally, in contrast to preceding reviews, we do not only consider positive impacts of explainability on other quality aspects, but we also take negative ones into account.

%Finally, preceding reviews mainly focused on positive relations between explainability and quality aspects while we also take negative ones in account.
\section{Research Goal and Design}
\label{sec:ResearchDesign}
\noindent We frame our study into the following three RQs:
\begin{framed}\noindent
	\textbf{RQ1:} What is a useful definition of explainability for the domains of software and requirements engineering?
	\newline \textbf{RQ2:} What are the quality aspects impacted by explainability in a system context?
	\newline \textbf{RQ3:} How does explainability impact these quality aspects?
\end{framed}
Since other disciplines have a long history working on explainability, their insights should prove valuable for software engineering and enable us to delineate the scope of the term explainability for this area. Accordingly, \textbf{RQ1} focuses on harnessing the work of other sciences in the field of explainability to compile a definition that is useful for the area of software and requirements engineering.

\textbf{RQ2} focuses on providing an overview of the quality aspects that may be impacted by explainability. Similar to the work of Leite and Capelli \cite{DoPrado2010}, who investigated the interaction between transparency and other qualities, our goal is to offer an overview for explainability and its impact on other quality aspects within a system.

With \textbf{RQ3} we want to assess what kind of impacts explainability has on other quality aspects. More specifically, our goal is to analyze the polarity of these impacts: whether they are positive or negative. To answer RQ2 and RQ3, we build a model and a catalogue that compiles knowledge about the impacts of explainability on other quality aspects.

\begin{figure}[htbp]
	\centering
	\includegraphics[width=0.48\textwidth]{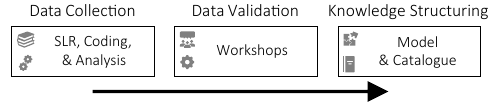}
	\caption{Overview of the research design}
	\label{fig:researchprocess}
\end{figure}

An overview of our research design is shown in \figurename~\ref{fig:researchprocess}.
Our research consisted of a multi-method approach that combined two qualitative methods to achieve higher reliability of our data. The first method focuses on systematic data collection and qualitative data analysis. For the data collection, we conducted an interdisciplinary \emph{SLR} that resulted in a total of 229 papers. We \emph{coded} the gathered data by using an open coding approach \cite{Saldana2021}. As a next step, we \emph{analyzed} the resulting codes for definitions of explainability (RQ1), for relationships between explainability and other quality aspects (RQ2), and for information about the polarity of these relationships (RQ3). To validate and complement our findings, we employed a second qualitative method: two \emph{workshops} with experts. Finally, we framed the obtained knowledge in a \emph{model} by structuring and grouping the quality aspects impacted by explainability along four dimensions and developed our \emph{catalogue} based on it.

The focus of this paper is on the results obtained through the qualitative research we conducted. For details on the SLR, especially the inclusion/exclusion criteria and the complete list of the papers analyzed through our literature review, please refer to our supplementary material \cite{AdditionalMaterial}. A more detailed description of the workshops can also be found there. Further results and details about the methodology employed in our literature review will be addressed in a future publication.

% \iffalse
% \begin{figure}[htbp]
% 	\centering
% 	\includegraphics[width=0.46\textwidth]{figs/Research-Procedure-1-v07.pdf}
% 	\caption{Overview of the Research Design}
% 	\label{fig:researchprocess}
% \end{figure}
% \fi

\subsection{Data Collection and Analysis}
%\subsection{Data Collection}
\subsubsection{Systematic Literature Review}
\label{sec:SLR}

We followed guidelines from Kitchenham et al. \cite{Kitchenham2007}, and Wohlin \cite{Wohlin2014} when conducting our SLR. The search strategy for our SLR consisted of a manual search followed by a snowballing process.

%First, we identified relevant venues for our manual search. After that, the selected venues were independently evaluated by invited researchers of the specific domains. We manually reviewed all of the selected journals and conference proceedings by judging their relevance according to our inclusion and exclusion criteria.
The manual search was performed independently by the authors of this paper and resulted in 104 papers. We used Fleiss' Kappa statistics \cite{Fleiss1971} to assess the reliability of the selection process. The calculated value of $\kappa = 0.81$ showed an almost perfect agreement \cite{Landis1977}. After the manual search, we performed snowballing to complement the search results. Our snowballing procedure included forward and backward snowballing \cite{Wohlin2014} and resulted in additional 125 papers. Overall, our SLR yielded a total of 229 papers. % A second snowballing iteration led to too many duplicates (> 75\%), which could have been caused by the large start set resulting from our manual search and first iteration.
The snowballing step was also independently conducted by the authors. The calculated value of $\kappa = 0.87$ also shows an almost perfect agreement.

This literature review process is partially based on a grounded theory (GT) approach for literature reviews proposed by Wolfswinkel et al. \cite{Wolfswinkel2013}. The goal of using this approach to reviewing the literature is to reach a detailed and relevant analysis of a topic, following some of the principles of GT. According to \cite{Wolfswinkel2013}, a literature review is never complete but at most saturated. This saturation is achieved when no new concepts or categories arise from the data. We followed this approach to decide when to conclude our snowballing process.

\subsubsection{Coding and Analysis}
\label{sec:AnAndCo}
%Each paper was conscientiously analyzed and the data appropriately extracted to answer the research questions. A data extraction form was used to ensure that this step was conducted in a consistent, complete, and precise manner. Subsequently, we coded our findings.

%Our data analysis process is based on data extracted from text excerpts of the selected papers. To extract this data, we analyzed whether the authors state an influence of explanations or explainability on a certain quality aspect of a system. If so, we selected the text excerpt and copied it into our database, along with information about the publication. To make sense of this data and extract the necessary information for our catalogue, we followed an open-coding approach~\cite{Saldana2021} for qualitative analysis.

We followed an open-coding approach~\cite{Saldana2021} for the qualitative analysis of the papers we found during our search. This approach consists of up to three consecutive coding cycles. For our first coding cycle, we applied \emph{Initial Coding} \cite{Charmaz2006} to preserve the views and perspectives of the authors in the code. In the second coding cycle, we clustered the initial codes based on similarities, using \emph{Pattern Coding}~\cite{Miles1994}. This allowed us to group the data from the first coding cycle into categories. Next, we discussed these categories until we reached an agreement on whether they adequately reflected the meaning behind the codes. These categories allowed us to structure the data for better analysis and to identify similarities.

For RQ2 and RQ3, we conducted a third coding cycle to further classify the categories into quality aspects. We applied \emph{Protocol Coding}~\cite{Boyatzis1998} as a procedural coding method in this cycle. For this method, we used a pre-established list of NFRs from Chung et al. \cite{Chung2012}. If any correspondence between a category and an NFR was found, we assigned the corresponding code. In the specific cases where we could not assign a corresponding NFR from \cite{Chung2012} to the data, we discussed together and selected a quality aspect that would adequately describe the idea presented in the text fragment. All coding processes were conducted independently by the authors of this paper. We had regular consensus sessions to discuss discrepancies. A list of all codes is available in our supplementary material \cite{AdditionalMaterial}.

\subsection{Data Validation}
\label{sec:Workshops}

We held two workshops to validate and augment the knowledge gathered during data collection: one with philosophers and psychologists, and one with requirements engineers. In both workshops, we discussed the categories and other relevant information that were identified during our coding. For RQ1, the categories consisted of competing definitions of explainability that we extracted from the literature. For RQ2, the categories consisted in the identified quality aspects that have a relationship with explainability. Finally, for RQ3, we identified the kind of impact that explainability can have on each of the extracted quality aspects.

\subsubsection{Workshop with Philosophers and Psychologists}
\label{sec:PhilWorkshop}
We validated the data related to RQ1 in a workshop with philosophers and psychologists (two professors, one postdoc, three doctoral candidates). Scholars in these disciplines have a long history in researching explanations and, thus, explainability. After consulting with experts from these disciplines on the workshop design, we decided on an open discussion. We instructed participants to hand in their notion of explainability prior to the meeting. During the workshop, we presented our coded data with the most prominent categories concerning RQ1. After a round of discussion, we presented the submitted definitions and compared them with our findings. We debated the similarities and differences between both and reached a consensus that eventually led to our proposed definition.

\subsubsection{Workshop with Requirements Engineers}
\label{sec:REWorkshop}
We validated the data related to RQ2 and RQ3 in a workshop with requirements engineers (three professors, two postdocs, one practitioner, one doctoral candidate). Two experts in the field of RE with experience in the topic of NFRs and software quality were consulted about the workshop design. The participants of this workshop also had to hand in a task in advance. For the task, we sketched scenarios where an explainable system was to be developed and sent the list of quality aspects we found, as seen in Fig.~\ref{fig:PolarityTable}. To avoid bias, we removed the polarities. Participants had to indicate which quality aspects were important in each system and what possible influence explainability could have on each of them. During the workshop, we discussed the outcomes of this assignment and had an open debate on the aspects on the list. Afterwards, we presented our findings and compared them with the received feedback. Overall, the experts were able to relate to each of the polarities we found. 

%\subsection{Knowledge Extraction and Structuring}
\subsection{Knowledge Structuring}
The last step of our research consisted of making sense of and structuring the knowledge collected in the previous stages.

\subsubsection{Framing the Results -- Model}
\begin{figure*}[!b]
	\centering
	\includegraphics[width=0.95\textwidth]{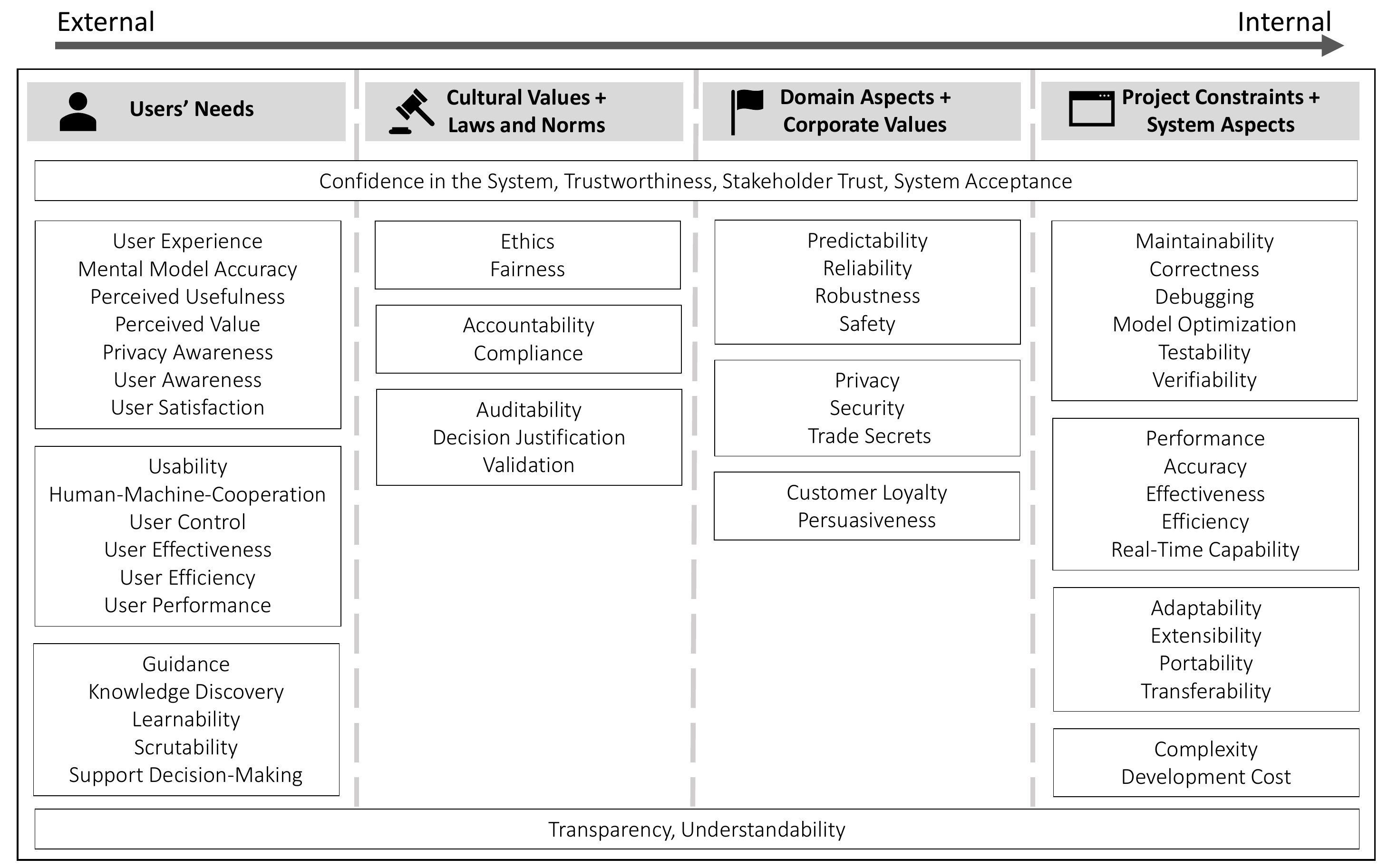}
	%\caption{A Model of the Impact of Explainability on Different Dimensions. Quality Aspects are grouped according to similarity.}
	%\caption{A Model of the quality aspects influenced by explainability, scattered across the four system dimensions}
	%\caption{How Explainability impacts Quality Aspects scattered along different Dimensions}
	\caption{A model illustrating the impact of explainability across different quality dimensions}
	%\caption{A model of the quality aspects impacted by explainability scattered across four system dimensions.}% The quality aspects are grouped according to similarity, based on our workshops' results.}
	\label{fig:Dimensions}
\end{figure*}

We built a model to frame our knowledge catalogue. This model illustrates the impact of explainability on several quality dimensions (see \figurename~\ref{fig:Dimensions}; RQ2). During the workshop with requirements engineers, we discussed possible ways to classify the different quality aspects. Here, the participants offered useful ideas. To further supplement these ideas, we consulted the literature and found three promising ways to classify the results. These three ways are analogous to the suggestions made by the workshop participants and supported us in the development of our model. 

\subsubsection{Catalogue Construction}
We summarized the results for RQ3 in a knowledge catalogue for explainability. Overall, we have extracted 57 quality aspects that might be influenced by explainability. We present these quality aspects and how they are influenced by explainability in~\figurename~\ref{fig:PolarityTable}. Additionally, we extracted a representative example from the literature for all positive and negative influences listed in our catalogue to show how this influence may come about. These examples also serve to illustrate our understanding of certain quality aspects.

We present the results for RQ1 in Sec.~\ref{sec:Definition} and the results for RQ2 and RQ3 in Sec.~\ref{sec:Model} and \ref{sec:Catalogue}.

\section{A Definition of Explainability}
\label{sec:Definition}

%Before requirements engineers can elicit the need for explainability in a system, they have to understand what explainability is in a system context. For this reason, finding a definition of explainability is the goal of our first RQ.

The domain of software engineering does not need a mere abstract definition of explainability, but one that focuses on requirements for explainable \emph{systems}. Before requirements engineers can elicit the need for explainability in a system, they have to understand what explainability is in a system context. For this reason, we provide a definition of what makes a system explainable to answer our first RQ.

Explainability is tied to disclosing information, which can be done by giving explanations. K{\"o}hl et al.~hold that what makes a system explainable is the access to explanations \cite{Koehl2019}. However, this leaves open what exactly is to be explained. In the literature, definitions of explainability vary considerably in this regard. Moreover, our review has revealed other aspects in which definitions of explainability differ. Consequently, there is not \emph{one} definition of explainability, but several complementary ones. Similarly, K{\"o}hl et al.~also found that there is not just one type of explainability, but that a system may be explainable in one respect but not in another \cite{Koehl2019}. Based on their definition of explainability, the definitions we found in the literature, and results from our workshop with philosophers and psychologists, we were able to develop an abstract definition of explainability that can be adjusted according to project or field of application.

\begin{framed}\noindent
 \textbf{Answering RQ1: }\emph{A system $S$ is explainable with respect to an aspect $X$ of $S$ relative to an addressee $A$ in context $C$ if and only if there is an entity $E$ (the explainer) who, by giving a corpus of information $I$ (the explanation of $X$), enables $A$ to understand $X$ of $S$ in $C$.}
\end{framed}

There were differences in the literature concerning the values of the following variables presented in the above definition: aspects of a system that should be explained, contexts in which to explain, the entity that does the explaining (the explainer), and addressees that receive the explanation. Being aware of these differences is crucial for requirement engineers, as they need to elicit the right kind of explainability for a project. We will shortly discuss some of these findings in what follows.

\paragraph{Aspects that should be explained} Concerning the aspects that should be explained, we found the following options in the literature and validated them during the workshop with philosophers and psychologists: the system in general \cite{Carvalho2019}, and, more specifically, its reasoning processes (e.g., inference processes for certain problems) \cite{Pieters2011}, its inner logic (e.g., relationships between the inputs and outputs) \cite{Koehl2019}, its model's internals (e.g., parameters and data structures) \cite{Holzinger2019}, its intention (e.g., pursued outcome of actions) \cite{Hois2019}, its behavior (e.g., real-world actions) \cite{Glass2008}, its decision (e.g., underlying criteria) \cite{Adadi2018}, its performance (e.g., predictive accuracy) \cite{Liao2020}, and its knowledge about the user or the world (e.g., user preferences) \cite{Glass2008}.

\paragraph{Contexts and Explainers} A context is set by a situation consisting of the interaction between a person, a system, a task, and an environment \cite{Dourish2004}. Plausible influences on the context are time-pressure, the stakes involved, and the type of system \cite{Langer2021}. Explainers refer to a system or specific parts of a system that supply its stakeholders with the needed information.

\paragraph{Addressee's Understanding} A vast number of papers in the literature make reference to the addressee's understanding as important factor for the success of explainability (e.g., \cite{Ribeiro2016, Rosenfeld2019, Arrieta2020, Carvalho2019, Miller2019}). Framing explainability in terms of under\-standing provides the benefit of making it measurable, as there are established methods of eliciting a person's understanding of something,  such as questionnaires or usability tests \cite{Chazette2020}.
\section{A Model of Explainability}
\label{sec:Model}

Models and catalogues compile knowledge about quality aspects and help to better visualize their possible impact on a system. Based on the data extracted from the literature and on our qualitative data analysis and validation, we were able to build a model and a catalogue for explainability. 
%The above definition is the first part of our model for explainability.
Overall, our model is divided into four dimensions. We considered three existing concepts to shape and compose these dimensions.

\begin{framed}\noindent
\textbf{Answering RQ2:} We framed the quality aspects that are impacted by explainability in a model that spans different quality dimensions of a system (Fig.~\ref{fig:Dimensions}).% Fig. \ref{fig:Dimensions} depicts the impact of explainability on quality aspects spread along four dimensions.
\end{framed}

The first concept connects to our definition and is based on the insight that understanding is pivotal for explainability. Langer et al.~tackle explainability from the perspective of the persons who inquire after explanations \cite{Langer2021}. Individuals differ in their background-knowledge, values, experiences, and many further respects. Accordingly, they also differ in what is required for them to understand certain aspects of a system. Furthermore, Langer et al.~also hold that some persons are more likely to be interested in a certain quality aspect than others. For instance, a developer may be more interested in the maintainability of a system than a user. They categorize quality aspects that are influenced by explainability according to so-called \emph{stakeholder classes} and distinguish the following ones: \emph{users}, \emph{developers}, \emph{affected parties}, \emph{deployers}, and \emph{regulators}. According to them, these classes should serve as a reference point when it comes to implementing explainability since the interests of different stakeholder classes may conflict \cite{Langer2021}. 

Chazette and Schneider identified six dimensions that affect the elicitation and analysis of explainability \cite{Chazette2020}: the \emph{needs and expectations of users}, \emph{cultural values}, \emph{corporate values}, \emph{laws and norms}, \emph{domain aspects}, and \emph{project constraints}. Their results indicate that different factors distributed across these dimensions influence the identification of explainability as being a necessary NFR within a system and the design choices towards its operationalization. We adopt and extend this notion in that we consider these dimensions to be decisive not only for the RE process but also for a system in general.

Finally, the external/internal quality concept based on the ISO 25010 \cite{ISO25010} and proposed by \cite{Freeman2009} is another way to categorize the quality aspects in our model. We consider the \emph{external} quality characteristics as the ones which are more related to the users or the \emph{quality in use}, and the \emph{internal} as the ones which are more related to the developers or the \emph{system itself}. As pointed out by McConnel \cite{Mcconnell2004}, the difference between internal and external characteristics is not completely clear-cut and affects several dimensions. Therefore, we do not assign clear-cut internal or external dimensions, but rather acknowledge a continuous shift from external to internal.

Based on these concepts, we developed a model of explainability and its impacts on other quality aspects. We frame the quality aspects along the four dimensions of our model: \emph{user's needs}, \emph{cultural values \& laws and norms}, \emph{domain aspects \& corporate values}, and \emph{project constraints \& system aspects}. Furthermore, we also identified quality aspects that are present in all dimensions. More details on the individual dimensions will be given in the next section, when we discuss the catalogue since the dimensions are closely linked to the quality aspects they frame. The dimensions and their respective quality aspects are illustrated in Fig.~\ref{fig:Dimensions}. In the figure, the quality aspects are grouped according to similarity, based on our workshops' results. Overall, the model should support requirements engineers in understanding how explainability can affect a system, facilitating requirements analysis.
\section{A Catalogue of Explainability's Impacts}
\label{sec:Catalogue}

\begin{figure*}[!ht]
	\centering
	\includegraphics[width=0.95\textwidth]{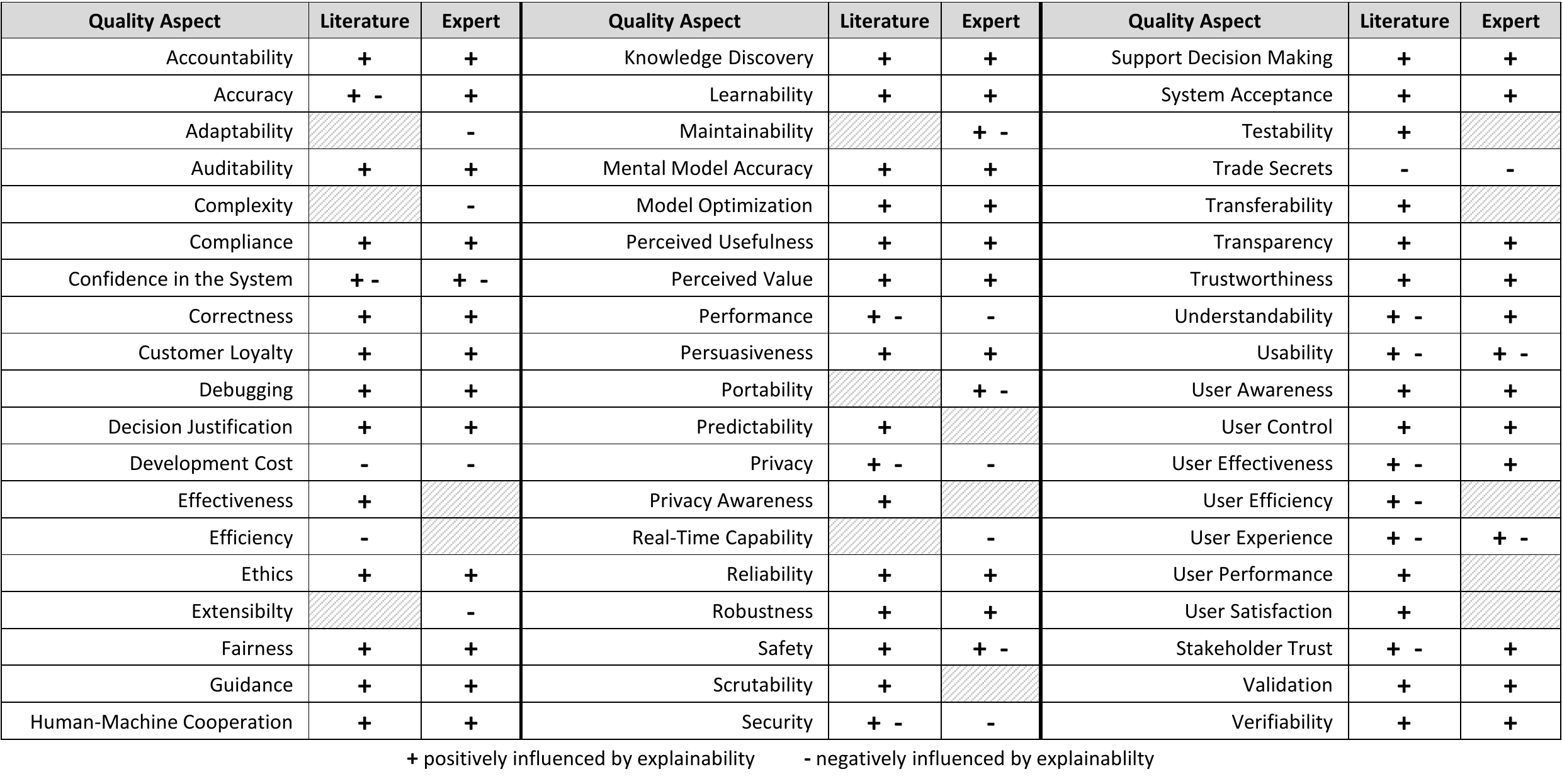}
	\caption{The knowledge catalog for explainability: how explainability impacts other quality aspects.}
	\label{fig:PolarityTable}
\end{figure*}

In what follows, we will present the catalogue and discuss the quality aspects in relation to our model. To this end, we will analyze them, whenever possible, based on the three categorizations we have described above: the stakeholders involved, the dimensions that affect the elicitation and analysis of explainability, and the external/internal categorization.

%\begin{framed}\noindent
%\textbf{Answering RQ3:}  Fig. \ref{fig:PolarityTable} lists all quality aspects found in our study and %the kind of impact explainability has on each one of these aspects according to our findings. 
%\end{framed}

%\begin{framed}\noindent
%\textbf{Answering RQ3:}  Based on our findings, we built a catalogue that lists all quality aspects and the kind of impact that explainability has on each one of these aspects (Fig. \ref{fig:PolarityTable}).
%\end{framed}

\begin{framed}\noindent
\textbf{Answering RQ3:} We built a catalogue that lists all quality aspects found in our study and the kind of impact that explainability has on each one of these aspects (Fig.~\ref{fig:PolarityTable}).
\end{framed}

%The symbols in \figurename~\ref{fig:PolarityTable} show whether the evidence found, either in the literature or during discussion in the workshop with requirements engineers, indicates a positive (+) or negative (-) influence of explainability on the quality aspect in question.

\subsection{Foundational Qualities}

Explainability can influence two quality aspects that have a crucial role: \textbf{transparency} and \textbf{understandability}. These quality aspects provide a foundation for all four dimensions, thereby having an influence on the other aspects inside these dimensions. Receiving explanations about a system, its processes and outputs can facilitate understanding on many levels \cite{Henin2019}. Furthermore, explanations contribute to a higher system transparency \cite{Chen2019}. For instance, understandability and transparency are required on a more external dimension so that users understand the outputs of a system, which may positively impact user experience. They are also important on a more internal dimension, where they can contribute to understanding aspects of the code, facilitating debugging and maintainability. 

%%%%%%%%%%%%%%%%%%%%%%%%%%%%%%%%%%%%%%%%%%%%%%%%%%%%%%%%%%%%%%%%%%%%%%%%%
\subsection{User's Needs}

Most papers concerning stakeholders in Explainable Artificial Intelligence (XAI) state \textit{users} as a common class of stakeholders (e.g., \cite{Arrieta2020, Preece2018, Weller2019}). This, in turn, also coincides with the view from requirements engineering, where (end) users also count as a common class of stakeholders \cite{Glinz2007a}. Among others, users take into account recommendations of AI systems to make decisions \cite{Hind2019}. Members of this stakeholder class can be medical doctors, loan officers, judges, or hiring managers. Usually, users are no experts regarding the technical details and the functioning of the systems they use \cite{Langer2021}.

When explainability is integrated into a system, different groups of users will certainly have different expectations, experiences, personal values, preferences, and needs. Such aspects mean that individuals can perceive quality differently. At the same time, explainability influences aspects that are extremely important from a user perspective.

The quality aspects we have associated with users are mostly external. In other words, they are not qualities that depend solely on the system. To be more precise, they depend on the expectations and the needs of the person who uses the system.

On a general level, the \textbf{user experience} can both profit and suffer from explainability. Explanations can foster a sense of familiarity with the system \cite{Riedl2019} and make it more engaging \cite{Mcinerney2018}. In this case, user experience profits from explainability. On the other side, explanations can cause emotions such as confusion, surprise \cite{Cai2019}, and distraction \cite{Miller2019}, harming the user experience. Furthermore, explainability has a positive impact on the \textbf{mental-model accuracy} of involved parties. By giving explanations, it is possible to make users aware of the system's limitations \cite{Liao2020}, helping them to develop better mental models of it \cite{Cai2019}. Explanations may also increase a user's ability to predict a decision and calibrate expectations with respect to what a system can or cannot do \cite{Liao2020}. This can be attributed to an improved \textbf{user awareness} about a situation or about the system \cite{Zhou2019}. Furthermore, explanations about data collection, use, and processing allow users to be aware of how the system handles their data. Thus, explainability may be a way to improve \textbf{privacy awareness} \cite{Arrieta2020}. Explainability can also positively impact the \textbf{perceived usefulness} of a system or a recommendation \cite{Zanker2012}, which contributes to the \textbf{perceived value} of a system, increasing users' perception of a system's competence \cite{Pu2006} and integrity \cite{Kizilcec2016} and leading to more positive attitudes towards the system \cite{Cramer2008}.
Finally, all of this shows that explainability can certainly positively impact \textbf{user satisfaction} with the system \cite{Cai2019}.

Explainability can also influence the \textbf{usability} of a system. On the positive side, explanations can increase the ease of use of a system \cite{Nunes2017}, lead to more efficient use \cite{Zhou2019}, and make it easier for users to find what they want \cite{Tintarev2007}. On the negative side, explanations can overwhelm users with excessive information \cite{Tsai2019} and can also impair the user interface design \cite{Chazette2020}. Explanations can help to improve \textbf{user performance} on problem solving and other tasks \cite{Kizilcec2016}. Another plausible positive impact of explainability is on \textbf{user effectiveness} \cite{Tintarev2012}. With explanations, users may experience greater accuracy in decision-making by understanding more about a recommended option or product \cite{Darlington2013}. However, user effectiveness can also suffer when explanations lead users to agree with incorrect system suggestions \cite{Bussone2015}. \textbf{User efficiency} is another quality aspect that can be positively and negatively influenced by explainability. Analyzing and understanding explanation takes time and effort \cite{Kumar2019}, possibly reducing user efficiency. Overall, however, the time needed to make a judgment could also be reduced with complementary information \cite{Tintarev2012}, increasing user efficiency. Furthermore, explanations may also give users a greater sense of \textbf{control}, since they understand the reasons behind decisions and can decide whether they accept an output or not \cite{Rosenfeld2019}. Explainability can also have a positive influence on \textbf{human-machine cooperation} \cite{Ribeiro2016} since explanations may provide a more effective interface for humans \cite{Dodge2019}, improving interactivity and cooperation \cite{Carvalho2020}, which can be especially advantageous in the case of cyber-physical systems.

Explainability can have a positive influence on \textbf{learnability}, allowing users to learn about how a system works or how to use a system \cite{Darlington2013}. It may also provide \textbf{guidance}, helping users in solving problems and educating them about product knowledge \cite{Putnam2019}. As these examples illustrate, explanations can \textbf{support decision-making} processes for users \cite{Nunes2017}. In some cases, this goes as far as enabling \textbf{scrutability} of a system, that is, enabling a user to provide feedback on a system's user model so that the system can give more valuable outputs or recommendations in the future \cite{Nunes2017}. Finally, explainability can help \textbf{knowledge discovery} \cite{Rosenfeld2019}. By making the decision patterns in a system comprehensible, knowledge about the corresponding patterns in the real world can be extracted. This can provide a valuable basis for scientific insight \cite{Liao2020}.

%%%%%%%%%%%%%%%%%%%%%%%%%%%%%%%%%%%%%%%%%%%%%%%%%%%%%%%%%%%%%%%%%%%%%%%%%
\subsection{Cultural Values \& Laws and Norms}

Although \cite{Chazette2020} distinguished \textit{Cultural Values} and \textit{Laws and Norms} as two separate dimensions and \cite{Langer2021} did the same for regulators and affected parties, we have combined them into one dimension because they are complementary and influence each other. The dimensions form a kind of symbiosis since, e.g., legal foundations are grounded, among others, on the basis of the cultural values of a society. We adopt the same approach for the dimensions discussed in Sec.~\ref{sec:Domain} and Sec.~\ref{sec:Project}.

Regulators commonly envision laws for people who could be affected by certain practices. In other words, regulators stipulate legal and ethical norms for the general use, deployment, and development of systems. This class of stakeholders occupies an extraordinary role, since they have a \denquote{watchdog} function concerning the systems and their use \cite{Langer2021}. Regulators can be ethicists, lawyers, and politicians, who must have the know-how to assess, control, and regulate the whole process of developing and using systems.

The restrictive measures by regulators are necessary, as the influence of systems is constantly growing and key decisions about people are increasingly automated -- often without their knowing \cite{Langer2021}. Affected parties are (groups of) people in the scope of a system's impact. They are stakeholders, as for them much hinges on the decision of a system. Patients, job or loan applicants, or defendants at court are typical examples of this stakeholder class \cite{Langer2021}.

In this dimension, cultural values represent the ethos of a society or group and influence the need for specific system qualities and how they should be operationalized \cite{Pacey1983, Kummer2012}. These values resonate in the conception of laws and norms, which enforce constraints that must be met and granted in the design of systems. Explainability can influence key aspects on this dimension.

With regard to the internal/external distinction, a clear attribution is not possible. Rather, the quality aspects seem to occupy a hybrid position. Whether or not they are present does not only depend on the system itself, but it also does not depend on a person using them. Rather, it depends on general conventions (e.g., legal, societal) that are in place. 

On the cultural side, explanations can contribute to the achievement of \textbf{ethical} decision-making \cite{Schneider2019} and, more specifically, ethical AI. On the one hand, explaining the agent's choice may support ensuring that ethical decisions are made \cite{Rosenfeld2019}. On the other hand, providing explanations can be seen as an ethical aspect itself. Furthermore, explainability may also contribute to \textbf{fairness}, enabling the identification of harms and decision biases to ensure fair decision-making \cite{Rosenfeld2019}, or helping to mitigate decision biases \cite{Liao2020}.

On the legal side, explainability can promote a system's \textbf{compliance} with regulatory and policy goals \cite{Gilpin2018b}. Explaining an agent's choice can ensure that legal decisions are made \cite{Rosenfeld2019}. A closely related aspect is \textbf{accountability}. We were able to identify a positive impact of explainability on this quality that occurs when explanations allow entities to be made accountable for a certain outcome \cite{Monteath2018}. In the literature, many authors refer to this as \textit{liability} \cite{Monteath2018} or \textit{legal accountability} \cite{Binns2018}.

In order to guarantee a system's adherence to cultural and legal norms, regulators and affected parties need several mechanisms that allow for inspecting systems. One NFR that can help in this regard is \textbf{auditability}. Explainability positively impacts this NFR, since explanations can help to identify whether a system made a mistake \cite{Bussone2015}, can help to understand the underlying technicalities and models \cite{Hois2019}, and allow users to inspect a system's inner workings to judge whether it is acceptable or not \cite{Mccarthy2004}. In a similar manner, \textbf{validation} can be positively impacted, since explainability makes it possible for users to validate system knowledge \cite{Darlington2013} or assess if a recommended alternative is truly adequate for them \cite{Nunes2017}. Exactly the latter aspect is essential for another quality that is helped by explainability, namely, \textbf{decision justification}. On the one hand, explanations are a perfect way to justify a decision \cite{Monteath2018}. On the other hand, they can also help to uncover whether a decision is actually justified \cite{Adadi2018}.

%%%%%%%%%%%%%%%%%%%%%%%%%%%%%%%%%%%%%%%%%%%%%%%%%%%%%%%%%%%%%%%%%%%%%%%%%
\subsection{Domain Aspects \& Corporate Values}
\label{sec:Domain}

People who decide where to employ certain systems (e.g., a hospital manager decides to bring a special kind of diagnosis system into use in her hospital) are deployers. Other possible stakeholders in this dimensions are specialists in the domain, known as domain experts. People have to work with the deployed systems and, consequently, new people fall inside the range of affected people \cite{Langer2021}.

This dimension is shaped by two aspects: 1) the corporate values and vision of an organization \cite{Thomsen2004}, and 2) the domain aspects that shape a system's design since explanations may be more urgent in some domains as in others.

We consider this dimension as more internal to the system, since it encompasses quality aspects that are more related to the domain or the values of the corporation or the team. Generally, the integration of such aspects affects the design of a system on an architectural level. However, there are some exceptions, as the organization's vision may aim at external factors like customer loyalty.

Explainability supports the \textbf{predictability} of a system by making it easier to predict a system's performance correctly and helping to determine when a system might make a mistake \cite{Lage2019}. Furthermore, explainability can support the \textbf{reliability} of a system \cite{Carvalho2019}. In general, explainability supports the development of more \textbf{robust} systems for critical domains \cite{Borgo2018}. All of this contributes to a positive impact on \textbf{safety}, helping to meet safety standards \cite{Rosenfeld2019}, or helping to create safer systems \cite{Guidotti2019}. On the negative side, explanations may also present safety risks by distracting users in critical situations.

Explanations are also seen as a means to bridge the gap between perceived \textbf{security} and actual security \cite{Pieters2011}, helping users to understand the actual mechanisms in systems and adapt their behavior accordingly. However, explanations may disclose information that makes the system vulnerable to attack and gaming \cite{Lepri2018}. Explainability can also influence \textbf{privacy} positively, since the principle of information disclosure can help users to discover what features are correlated with sensitive information that can be removed \cite{Hohman2019}. By the same principle, however, privacy can be hurt since one may need to disclose sensitive information that could jeopardize privacy \cite{Zhou2019}. Explainability can also threaten model confidentiality and \textbf{trade secrets}, which companies are reluctant to reveal \cite{Arrieta2020}.

Explainability can contribute to \textbf{persuasiveness}, since explanations may increase the acceptance of a system's decisions and the probability that users adopt its recommendations \cite{Nunes2017}. Furthermore, explainability influences \textbf{customer loyalty} positively, since it supports the continuity of use \cite{Riedl2019} and may inspire feelings of loyalty towards the system \cite{Tintarev2007}.

%%%%%%%%%%%%%%%%%%%%%%%%%%%%%%%%%%%%%%%%%%%%%%%%%%%%%%%%%%%%%%%%%%%%%%%%%
\subsection{Project Constraints \& System Aspects}
\label{sec:Project}

Individuals who design, build, and program systems are, among others, developers, quality engineers, and software architects. They count as stakeholders, as without them the systems would not exist in the first place. Generally, representatives of this group have a high expertise concerning the systems and a strong interest in creating and improving them.

This dimension is shaped by two aspects: project constraints and system aspects. The project constraints are the non-technical aspects of a system \cite{Carvallo2006}, while system aspects are more related to internal aspects of the system, such as performance and maintainability.

The quality aspects framed in this dimension are almost entirely internal in the classical sense, since they correspond to the most internal aspects of a system or the process through which the system is built.

Explainability can have both a positive and negative impact on \textbf{maintainability}. On the one hand, it can facilitate software maintenance and evolution by giving information about models and system logic. On the other hand, the ability to generate explanations requires new components in a system, hampering maintenance. A positive impact on \textbf{verifiability} was also identified, when explanations can work as a means to ensure the correctness of the knowledge base \cite{Darlington2013} or to help users evaluate the accuracy of a system's prediction \cite{Zhou2019b}. \textbf{Testability} falls in the same line, since explanations can help to evaluate or test a system or a model \cite{Rosenfeld2019}. Explainability has a positive influence on \textbf{debugging}, as explanations can help developers to identify and fix bugs \cite{Adadi2018}. Specifically, in the case of ML applications, this could enable developers to identify and fix biases in the learned model and, thus, \textbf{model optimization} is positively affected \cite{Mathews2019}. Overall, all these factors can help increase the \textbf{correctness} of a system, by helping to correct errors in the system or in model input data \cite{Monteath2018}.

%% Performance cluster
The overall \textbf{performance} of a system can be affected both positively and negatively by explainability. On the one hand, explanations can positively influence the performance of a system by helping developers to improve the system \cite{Ribeiro2016}. In this regard, explainability positively influences system \textbf{effectiveness}. On the other hand, however, explanations can also lead to drawbacks in terms of performance \cite{Kumar2019} by requiring loading time, memory, and computational cost \cite{Chazette2020}. Thus, as the additional explainability capacities are likely to require computational resources, the \textbf{efficiency} of the system might decrease \cite{Adadi2018}. Another quality that is impacted by explainability is \textbf{accuracy}. For instance, in the ML domain, the accuracy of models can benefit from explainability through model optimization \cite{Mathews2019}. On the negative side, there exists a trade-off between the predictive accuracy of a model and explainability \cite{Adadi2018}. A system that is inherently explainable, for instance, may have to sacrifice predictive power in order to be so \cite{Holzinger2019}. Explainability may have a negative impact on \textbf{real-time capability} since the implementation of explanations could require more computing power and additional processes, such as logging data, might be involved.  

%% Adaptability, Extensibility, Portability, and Transferability
\textbf{Adaptability} can be negatively impacted, for example, if lending regulations in a financial software have changed and an explanation module in the software is also affected. Next, assume that a new module should be added to a system. The quality aspect involved here is \textbf{extensibility}, which in turn is negatively impacted by explainability. Merely adding the new module is already laborious. If the explainability is also affected by this new module, the required effort increases again. Depending on the architecture of the software, it may even be impossible to preserve the system's explainability. Explanations affect the \textbf{portability} of a system as well. On the negative side, an explanation component might not be ported directly because it uses visual explanations, but the environment to which system is to be ported to has no elements that allow for visual outputs. On the positive side, explainability helps \textbf{transferability} \cite{Chen2018}. Transferability is the possibility to transfer a learned model from one context to another (thus, it can be seen as a special case of portability for ML applications). Explanations may help in this regard by making it possible to identify the context from and to which the model can be transferred \cite{Chen2018}.

Overall, the inclusion of explanation modules can increase the \textbf{complexity} of the system and its code, influencing many of the previously seen quality aspects. In particular, as an explainability component needs additional development effort and time, it can result in higher \textbf{development costs} \cite{Koehl2019}. 

%%%%%%%%%%%%%%%%%%%%%%%%%%%%%%%%%%%%%%%%%%%%%%%%%%%%%%%%%%%%%%%%%%%%%%%%%
\subsection{Superordinated Qualities}

We were able to identify some aspects that hold regardless of dimension. These aspects are commonly seen as some kind of superordinated goals of explainability. For instance, organizations and regulators have been lately focusing on defining core principles (or "pillars") for responsible or \textbf{trustworthy} AI. Explainability has been often listed as one of these pillars \cite{Arrieta2020}. Overall, many of the quality aspects we could find in the literature contribute to trustworthiness. For instance, explanations can help to identify whether a system is safe and whether it complies to legal or cultural norms. Ideally, \textbf{confidence} and \textbf{trust} in a system originate solely from trustworthy systems. Although one could trust an untrustworthy system, this trust would be unjustified and inadequate. For this reason, explainability can both contribute to and hurt trust or confidence in a system \cite{Pieters2011, Zhou2019}. Regardless of the system's actual trustworthiness, bad explanations can always degrade trust \cite{Pieters2011}. Finally, all of this can influence the system's \textbf{acceptance}. A system that is trustworthy can gain acceptance \cite{Glass2008} and explainability is key to this.
\section{Discussion}
\label{sec:Discussion}

Explainability is a new NFR that echoes the demand for more human oversight of systems \cite{Thiebes2020}. It can bring positive or negative consequences across all quality dimensions: from users' needs to system aspects. Explainability's impact on so many crucial dimensions illustrates the growing need to take explainability into account while designing a system. Currently, however, the RE community still lacks guidance on how to do so. Building appropriate elicitation techniques and developing adequate tools to capture explainability requirements are challenges that still need to be addressed.

To this end, our first contribution is a helpful definition of explainability for software and requirements engineers (Sec.~\ref{sec:Definition}). This definition points out what should be considered when dealing with requirements and the appropriate functionality for explainable systems: aspects that should be explained, contexts, explainers, and addressees. Being aware of these variables facilitates the software development process, supporting the elicitation and specification of explainability requirements. In this sense, the possible values (e.g., reasoning process) we found in the literature can serve as an abstract starting point during requirements analysis. Overall, our definition can serve as a template to help engineering explainable systems and to make good design choices towards explainability requirements.

In contrast, poor design choices regarding explainability can negatively affect the relationship with the user (e.g., user experience issues), interfere with important quality aspects for a corporation (e.g., damaging brand image and customer loyalty), and bring disadvantages for the project or the system (e.g., increasing development costs or hindering system performance). This kind of impact may stem from the fact that explainability might be seen as an aspect of communication between systems and humans. Depending on how it happens in practice, communication can either strengthen or harm relationships. %We may apply the same principle to explainability.

Research in RE can profit from insights of other disciplines when it comes to explainability. The fields of philosophy, psychology, and HCI, for example, have long researched aspects such as explanations or human interaction with systems (see \cite{Miller2019} for research concerning explanations in several disciplines). At the same time, requirements engineers can contribute to the field of explainability by studying how to include such aspects in systems and adapt development processes. This knowledge, scattered among different areas of knowledge, must be made available and integrated into the development of systems.

To this end, additional contributions of this paper are a model (Sec.~\ref{sec:Model}) and a knowledge catalogue (Sec.~\ref{sec:Catalogue}). Conceptual models are useful to abstract, comprehend, and communicate information. Among others, our catalogue can serve as checklist during elicitation and also during trade-off analysis. It can help software engineers avoid conflicts between quality aspects and choose the best strategies for achieving the desired quality outcomes. Both artifacts contain information that may be used to turn explainability into a positive catalyst for other essential system qualities in modern systems.

On a general level, building these artifacts has revealed that there is much to do in the field of NFRs. On the one hand, we believe that there may be other emerging NFRs besides explainability. Aspects such as human-machine cooperation, privacy awareness, and mental model accuracy show that there are specific needs that should be better understood when developing modern systems. Furthermore, ethics, fairness, and legal compliance are all good examples of quality aspects that are gaining in importance and should be better researched \cite{Aydemir2018}.

On the other hand, we have identified that explainability can exhibit an impact on nearly all traditional NFRs that can be found in the ISO 25010~\cite{ISO25010}: performance, efficiency, usability, reliability, security, maintainability, and portability. As such, the importance of explainability has to be further acknowledged. In this line of thought, the impact of other NFRs on explainability should be better researched and existing catalogues could be updated to incorporate explainability. The RE community needs to explore what kind of activities, methods, and tools need to be incorporated into the software development process in order to accommodate the necessary steps towards building explainable systems. Our work is an essential step in this direction.
\section{Limitations and Threats to Validity}
\label{sec:ThreatsToValidity}
%This section discusses the limitations of the assertions made based on the literature review, coding procedure, and workshops. They are presented jointly because they complement each other.
Our work is exclusively based on qualitative data analysis. Consequently, there is the possibility that the results are affected by subjectivity during analysis. Therefore, we decided on a multi-method approach to produce results that are more robust and compelling than single method studies. Next, we discuss the main threats to validity in each part of our research.

\paragraph{SLR and coding} The review process assumed a common understanding among all researchers involved in this work with respect to the search and analysis methods used. Results could be subject to bias if the methods and concepts are misunderstood. We mitigated this threat by elaborating a review protocol and discussing it before starting the review to reach a good level of shared understanding. We have formulated inclusion and exclusion criteria to reduce biases due to subjective decisions in the selection process. Some criteria, such as the publication period, are objective, while others, focusing on the content of the papers, are still subjective. To decrease the amount of researcher bias, we conducted the analysis independently. For both the literature review and the coding process, in case of disagreement, the decision on inclusion or exclusion (for a paper) or the code assignment (for the extracted data) was taken by all researchers and validated by the Fleiss' Kappa statistic. %As a further step to eliminate researchers' bias, we validated our final findings by means of two workshops with experts from the domains of RE, psychology, and philosophy. 

\paragraph{Explainability catalogue} The clustering and categorization of the quality aspects into their different dimensions was prone to subjective judgment. As steps to mitigate this, we rooted this categorization on well-known concepts present in the literature and conducted workshops with experts. This allowed us to inspect our clustering through internal and external reviews. During the internal reviews, the categorization was discussed among the authors to clarify ambiguities and reach agreements. During the external reviews, we compared the findings from the literature with expert knowledge. Due to these review processes, we are confident to have achieved a proper level of validity of the catalogue. Moreover, as researchers we are confident that both our catalogue and model are reasonably accurate for the field studied, developed over debates that formed our shared knowledge on the subject. 
\section{Conclusion and Future Work}
\label{sec:Conclusion}

Explainability is increasingly seen as an appropriate means of achieving essential quality aspects in a system, such as transparency, accountability, and trustworthiness. As building these values into our systems becomes more urgent, there is a need for tools and methods that help elicit, implement, and validate related requirements. For this reason, we should be concerned with understanding explainability as a whole: its meaning, its effects, its taxonomy. 

In this sense, our proposed definition can help to facilitate communication and align expectations when referring to explainability. Our model can help professionals to understand its taxonomy, and our catalogue can help to identify conflicts between explainability and other important qualities. In holding this knowledge, it is possible to think of design strategies and implementation level solutions that result in positive effects for the stakeholders involved.

As a next step, we want to create a quality model that structures and expands the gathered knowledge with specific characteristics and aspects of explanations. Furthermore, we need to investigate what kinds of explainability-related activities should be integrated into the software development process to successfully develop explainable systems. Overall, we hope that our work lays the foundation for the RE community to better understand and investigate the topic of explainability.
\section*{Acknowledgments} 
\label{sec:ack}
\addcontentsline{toc}{section}{Acknowledgments}
This work was supported by the research initiative Mobilise between the Technical University of Braunschweig and Leibniz University Hannover, funded by the Ministry for Science and Culture of Lower Saxony and by the Deutsche Forschungsgemeinschaft (DFG, German Research Foundation) under Germany’s Excellence Strategy within the Cluster of Excellence PhoenixD (EXC 2122, Project ID 390833453). Work on this paper was also funded by  the Volkswagen Foundation grant AZ 98514 \href{https://explainable-intelligent.systems}{\enquote{Explainable Intelligent Systems}} (EIS) and by the DFG grant 389792660 as part of \href{https://perspicuous-computing.science}{TRR~248}. We thank Martin Glinz for his feedback on our research design. Furthermore, we thank all workshop participants, the anonymous reviewers, and the colleagues who gave feedback on our manuscript.

%We thank Andreas Vogelsang, Anna Averbakh, Anne Hess, Daniel Oster, Eric Knauss, Eva Schmidt, Kevin Baum, Kurt Schneider, Lena Kästner, Markus Langer, Nils Prenner, Sarah Sterz, and Verena Klös for participating in our workshops. We furthermore thank Dieter Belle.

%\cleardoublepage
%\interlinepenalty=10000
%\clearpage 
%\raggedbottom
\bibliographystyle{IEEEtran}
\bibliography{IEEEabrv,bibtex/bib/biblio}

\end{document}